\documentclass[journal=nalefd,manuscript=letter]{achemso}
\usepackage{chemformula} 
\usepackage[T1]{fontenc} 
\mciteErrorOnUnknownfalse
\usepackage{graphicx}
\graphicspath{ {./images/} }
\usepackage{siunitx}

\author{David Jaeger} 
\affiliation{Department of Physics, University of Basel, 4056 Basel, Switzerland} 
\alsoaffiliation{Swiss Nanoscience Institute, University of Basel, 4056 Basel, Switzerland}

\author{Francesco Fogliano} 
\affiliation{Department of Physics,  University of Basel, 4056 Basel, Switzerland}

\author{Thibaud Ruelle} 
\affiliation{Department of Physics, University of Basel, 4056 Basel, Switzerland}

\author{Aris Lafranca} 
\affiliation{Swiss Nanoscience Institute, University of Basel, 4056 Basel, Switzerland}

\author{Floris Braakman} 
\affiliation{Department of Physics, University of Basel, 4056 Basel, Switzerland}

\author{Martino Poggio} 
\affiliation{Department of Physics, University of Basel, 4056 Basel, Switzerland} 
\alsoaffiliation{Swiss Nanoscience Institute, University of Basel, 4056 Basel, Switzerland}
\email{martino.poggio@unibas.ch}

\title[Mechanical mode imaging of a high-Q hybrid hBN/\ch{Si3N4} resonator]
  {Mechanical mode imaging of a high-Q hybrid hBN/\ch{Si3N4} resonator}

\begin{document}







\begin{abstract}
We image and characterize the mechanical modes of a 2D drum resonator made of hBN suspended over a high-stress \ch{Si3N4} membrane. Our measurements demonstrate hybridization between various modes of the hBN resonator and those of the \ch{Si3N4} membrane. The measured resonance frequencies and spatial profiles of the modes are consistent with finite-element simulations based on an idealized geometry. Spectra of the thermal motion reveal that, depending on the degree of hybridization with modes of the heavier and higher-quality-factor \ch{Si3N4} membrane, the quality factors and the motional mass of the hBN drum modes can be shifted by orders of magnitude. This effect could be exploited to engineer hybrid drum/membrane modes that combine the low motional mass of 2D materials with the high quality factor of \ch{Si3N4} membranes for optomechanical or sensing applications.
\end{abstract}


\section{Main Text}
2D materials enable the fabrication of nanomechanical resonators with high aspect ratios, very 
low mass, and high resonance frequencies \cite{bunchElectromechanicalResonatorsGraphene2007,
zhengHexagonalBoronNitride2017,castellanos-gomezandresSingleLayerMoS22013, steenekenDynamics2DMaterial2021}. 
They have remarkable mechanical properties, such as extremely high fracture strength and Young's Modulus \cite{leeMeasurementElasticProperties2008}. 
A wide variety of 2D materials are now subject of large research efforts, due to their unique electronic \cite{chenPerformanceMonolayerGraphene2009,chenGrapheneMechanicalOscillators2013}, magnetic \cite{gongDiscoveryIntrinsicFerromagnetism2017a,huangLayerdependentFerromagnetismVan2017a} and optical \cite{splendianiEmergingPhotoluminescenceMonolayer2010,srivastavaOpticallyActiveQuantum2015} features, as well as the potential to combine these materials in layered heterostructures \cite{novoselov2DMaterialsVan2016a}. 
Hexagonal boron-nitride (hBN) is known to have comparable mechanical properties to 
graphene \cite{falinMechanicalPropertiesAtomically2017}, but in contrast to graphene and other 2D materials it has a large bandgap of almost \SI{6}{\eV} \cite{watanabeDirectbandgapPropertiesEvidence2004}, making it a transparent insulator with low absorption in the visible and near infra-red range \cite{leeRefractiveIndexDispersion2019}.
hBN has also been shown to host bright and stable quantum emitters \cite{tranQuantumEmissionHexagonal2016}, which are strain-coupled to the motion of the crystal lattice \cite{mendelsonStrainInducedModificationOptical2020, lazicDynamicallyTunedNonclassical2019}. 
These properties make hBN a prime candidate for optomechanical devices and integration into high-finesse optical cavities \cite{shandilyaHexagonalBoronNitride2019}. 

Most devices based on suspended flakes of 2D materials are fabricated by direct exfoliation \cite{bunchElectromechanicalResonatorsGraphene2007, leeHighFrequencyMoS22013} or dry stamping onto a pre-patterned substrate \cite{castellanos-gomezDeterministicTransferTwodimensional2014}.
Due to the thickness of the substrate and the fact that there is usually no optical access 
from both sides, these samples are not suitable for integration with micro-scale optical cavities in the membrane-in-the-middle (MIM) configuration \cite{ruelleTunableFiberFabry2022, flowers-jacobsFibercavitybasedOptomechanicalDevice2012a, rochauDynamicalBackactionUltrahighFinesse2021a}.
In addition, mechanical mode imaging on such devices has revealed that these transfer
techniques have a deleterious effect on the mode shapes and potentially on the mechanical 
properties of the resonators, due to the inhomogeneous stress they impart to the flake \cite{zhengHexagonalBoronNitride2017,
davidovikjVisualizingMotionGraphene2016,
garcia-sanchezImagingMechanicalVibrations2008}.
Aside from degrading the overall performance of the devices and their reproducibility, this is also problematic for sensing implementations that require calibrated mode shapes \cite{davidovikjVisualizingMotionGraphene2016}. 
Suspended devices made from 2D materials grown by chemical vapour deposition are fabricated with wet transfer techniques \cite{reinaLargeAreaFewLayer2009}, but they usually exhibit many folds, cracks or other imperfections that affect the resulting mechanical resonators to a similar degree \cite{zandeLargeScaleArraysSingleLayer2010,cartamil-buenoMechanicalCharacterizationCleaning2017}.  
	
\begin{figure*}[ht]
  \includegraphics[width=\textwidth]{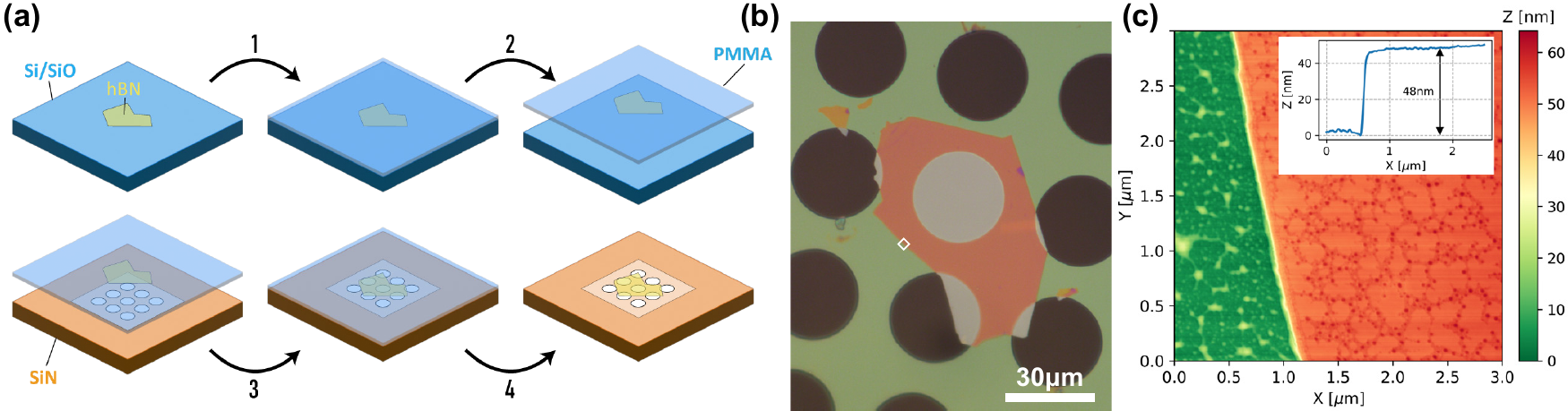}
  \caption{(a) Sample fabrication procedure. hBN flakes are exfoliated onto Si/\ch{SiO2} substrates and then spin-coated with PMMA in step 1. In step 2, the oxide layer is removed, releasing the PMMA layer with embedded hBN flakes. In step 3, this PMMA layer is brought into contact with the \ch{Si3N4} membrane, producing a hBN drum resonator. Finally, in step 4, the PMMA is removed in a solvent clean. (b) Photograph of a finished sample (c) AFM measurement of the area indicated by the small white rectangle in (b), giving a flake thickness of \SI{48}{\nm}.}
  \label{fgr:samplefab}
\end{figure*}

In this work, we employ a wet transfer technique to fabricate devices from exfoliated flakes of hBN (Fig. \ref{fgr:samplefab} (a)) which are placed on top of holes in \ch{Si3N4} membranes, resulting in thin devices with optical access from both sides (Fig. \ref{fgr:samplefab} (b)). 
We characterize such a hBN/\ch{Si3N4} mechanical resonator by measuring thermal mode spectra, as well as detailed spatial mode shapes. 
We find that the mode shapes match well with COMSOL Multiphysics simulations based on idealized geometries.
Finally, we explore how the hybridization between the mechanical modes of the \ch{Si3N4} membrane and of the hBN drum affects the mechanical properties of the latter.

While hybridization between 2D material resonators and underlying \ch{Si3N4} membranes has been observed before \cite{schwarzDeviationNormalMode2016,
singhGiantTunableMechanical2020}, here we focus on the changes in the quality factor (Q) and 
effective mass ($m^{*}$). 
In our case, the underlying \ch{Si3N4} membrane has a much higher Q than the hBN drum.
This results in hybridization where the \ch{Si3N4} membrane lends its mechanical properties to the modes of the hBN drum, which could be useful for sensing applications \cite{reicheIntroductionCoresonantDetection2015} and the engineering of functionalized mechanical systems.

Our device fabrication starts with exfoliated hBN flakes (hq-graphene) on a Si/\ch{SiO2} substrate with a \SI{300}{\nm} oxide layer.
hBN offers a poor optical contrast on PDMS, which is typically used in stamping techniques \cite{zhengHexagonalBoronNitride2015}. 
However, on a Si/\ch{SiO2} substrate the flakes can be carefully evaluated and selected, making it possible to find flakes without defects, steps or folds.
The Si/\ch{SiO2} chip is then spin-coated with a layer of Poly(methyl methacrylate) (PMMA), after which the oxide layer is removed in a 2M \ch{NaOH} solution.
This results in a floating PMMA membrane with hBN flakes attached to its bottom surface (see Fig. \ref{fgr:samplefab} (a) step 1-2).
We have found that a slow etching of the \ch{SiO2} at room temperature yields a smoother and cleaner PMMA membrane compared to a faster process at elevated temperatures, as is often employed.
After replacing the NaOH solution with DI water in several rinsing steps, the PMMA/hBN membrane is transferred to another vessel where it floats above the target substrate.
This substrate is a  stoichiometric (\SI{900}{\MPa} stress) \ch{Si3N4} membrane (Norcada). 
It is \SI{200}{\nm} thick, has lateral dimensions of \SI{300}{\um} x \SI{300}{\um}, and features \SI{30}{\um} diameter holes in a grid pattern.
The \ch{Si3N4} membrane is glued with Crystalbond to a metal holder so as to not float away or move during the transfer.
To position the PMMA membrane on top of the \ch{Si3N4} chip, we use a micromanipulator setup under an optical microscope.
While positioning the hBN flake above the target hole in the \ch{Si3N4} membrane, the water is slowly removed until the hBN flake settles on top of the hole (Fig. \ref{fgr:samplefab} (a) step 3).
The device is left to dry over night, after which the PMMA is removed in acetone followed by IPA at \SI{50}{\celsius} for \SI{30}{\min} each (Fig. \ref{fgr:samplefab} (a) step 4). 
Finally, the sample is put into a UV/Ozone cleaner for \SI{10}{\min} to remove organic residue.

\begin{figure*}[ht]
  \includegraphics[width=\textwidth]{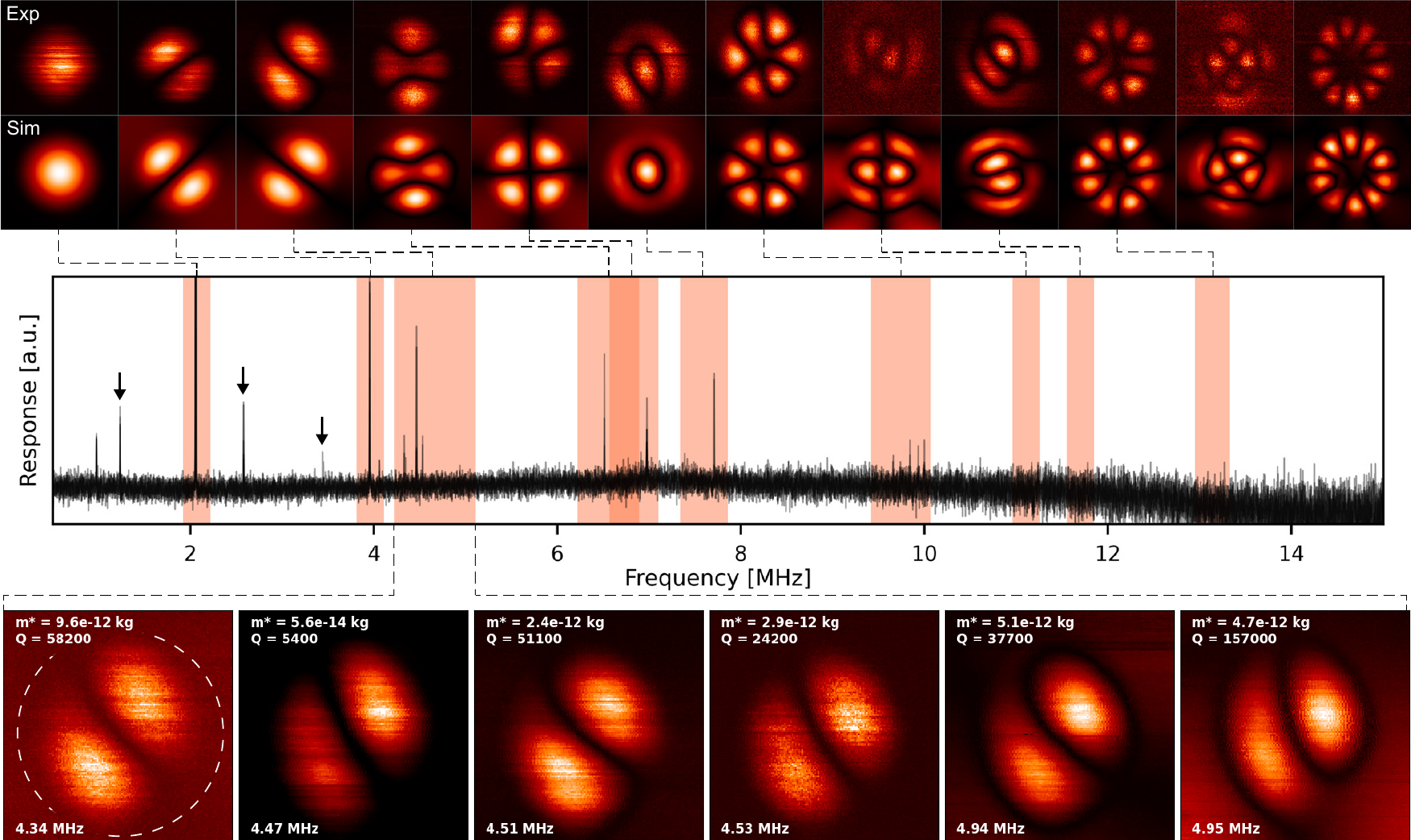}
  \caption{Mode images and spectrum obtained for the sample shown in Fig. \ref{fgr:samplefab} (b). Top row shows the measured mode images while the second row shows mode images simulated with COMSOL. Since often more than one mode could be found with matching mode images, the modes are represented with a shaded area rather than with a specific peak in the thermal spectrum shown below.  The last two mode images are beyond the frequency range of the thermal spectrum shown, therefore no shaded area is indicated. Arrows highlight modes of the \ch{Si3N4} membrane that did not hybridize with any hBN drum modes, starting with the fundamental \ch{Si3N4} mode. The last row shows an example of a series of apparently identical modes in one of the shaded areas. $m^{*}$, Q and resonance frequency are displayed on each of the six graphs.}
  \label{fgr:overviewmodes}
\end{figure*}

The characterization of the sample is carried out in a room temperature vacuum chamber (pressure <\SI{d-4}{\milli\bar}) with optical access for a modified Michelson interferometer (see Ref.~\citenum{bargMeasuringImagingNanomechanical2017}).
A single-frequency diode laser (\SI{630}{\nm}) is used to detect the motion of the hBN drum in a balanced detection scheme.
The length of the reference arm of the interferometer is stabilized via a feedback loop to keep the interferometer stable and at maximum sensitivity throughout the measurements.
The mechanical response is measured by exciting the sample resonantly with a piezoelectric shaker to an amplitude of a few $nm$, while recording the demodulated signal with a lock-in amplifier (Zurich HF2LI).
A stack of piezo scanners (Attocube ANSxy50) makes it possible to map the reflected intensity of the sample, as well as to image the mechanical modes as in Fig. \ref{fgr:overviewmodes}.

Mode imaging is not only a tool that can give insight into the imperfections (folds, wrinkles, inhomogeneous strain \latin{etc.}) of the membrane and their effect on the mode shapes, but it also allows to identify to which mode each resonance peak found in the spectrum belongs. 
This turns out to be particularly useful when the resonance spectrum is not well predicted by theory, or as in our case, is complicated by hybridization of the modes.

\begin{figure*}[ht]
  \includegraphics[width=\textwidth]{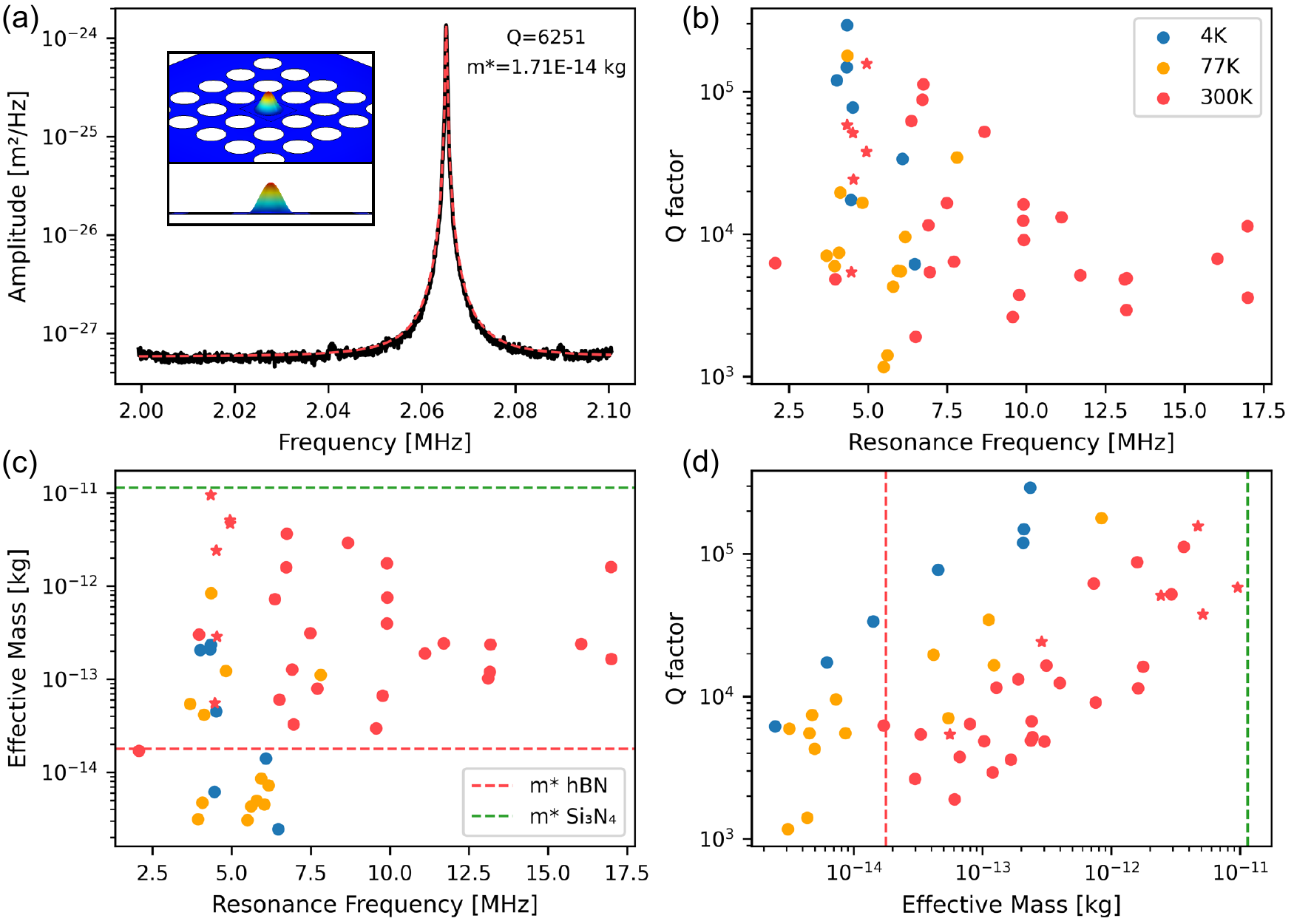}
  \caption{(a) Thermal spectrum of the fundamental mode with fit in red. Q and $m^{*}$ extracted from this fit are shown to the right of the peak, while the inset to the left shows the expected deflection as simulated with COMSOL. The motion is confined to the hBN that covers the central hole. (b),(c) Q and $m^{*}$ as a function of the resonance frequency for all observed modes at different temperatures. The six modes shown at the bottom of Fig. \ref{fgr:overviewmodes} are represented with star shaped markers. (d) Q vs. $m^{*}$. The dashed lines in (c),(d) indicate $m^{*}_{th}$ of hBN and \ch{Si3N4}.}
  \label{fgr:QandMass}
\end{figure*}

We have characterized and imaged a wide array of mechanical modes visible in the thermal spectrum (Fig. \ref{fgr:overviewmodes}) of the sample shown in Fig. \ref{fgr:samplefab} (b).
This data gives insight into the resonance spectrum and confirms that the mode shapes (first row of images in Fig. \ref{fgr:overviewmodes}) match the ones simulated with COMSOL (second row of images in Fig. \ref{fgr:overviewmodes}). 
We can also use the thermal spectra to characterize the mechanical properties of the modes in question, in particular their $m^{*}$ and Q.
For example, the thermal spectrum recorded for the fundamental mode of the hBN drum at around \SI{2}{\MHz} is shown in Fig. \ref{fgr:QandMass} (a).
The theoretical effective mass $m^{*}_{th} = \SI{1.89e-14}{\kg}$ agrees with the measured $m^{*} = \SI{1.71e-14}{\kg}$. 
For the theoretical value, we assumed a density of hBN of $\rho_{hBN}=\SI{2100}{\kg\per\m\cubed}$ and a ratio between $m^{*}$ and geometrical mass $m$ of  $m^{*}/m=0.265$. \cite{hauerGeneralProcedureThermomechanical2013a}.
The drum has a diameter of $d=\SI{30}{\um}$ and a thickness of $t=\SI{48}{\nm}$ (Fig. \ref{fgr:samplefab} (c)).
To fit the thermal spectra we use the power spectral density
\begin{equation}
S_{xx}(\omega)=\frac{2\Gamma k_B T}{m^{*}((\omega_{0}^2-\omega^2)^2+\Gamma^2 \omega^2)}
	\label{eqn:effmass}
\end{equation}
where $\omega /2\pi$ is the frequency ($\omega_0$ at resonance), $\Gamma=\omega_{0}/Q$ the linewidth, $T$ the temperature, and $k_B$ the Boltzmann constant.
We assume that the sample is thermalized with the surrounding bath. 
The optical power ($\sim$ \SI{100}{\uW}) was chosen so as to cause a negligible shift in the resonance frequencies of the hBN drum. 

We measure a quality factor $Q=6250$ for the fundamental mode of the hBN drum. 
We are not aware of a device with higher Q at room-temperature based on a 2D material.
Typically, Q-factors of such devices are in the range of $\num{e1}-\num{e2}$ \cite{vanleeuwenTimedomainResponseAtomically2014,
leeHighFrequencyMoS22013,
bunchElectromechanicalResonatorsGraphene2007,
castellanos-gomezandresSingleLayerMoS22013,
kramerStraindependentDampingNanomechanical2015,
cartamil-buenoMechanicalCharacterizationCleaning2017}.

We often find several copies of a specific mode shape in a frequency interval indicated by the shaded regions in Fig. \ref{fgr:overviewmodes}. 
An example of a series of such modes with observable thermal motion can be seen in the bottom of Fig. \ref{fgr:overviewmodes}.
We attribute the presence of such copies to hybridization between an hBN drum mode and several modes of the \ch{Si3N4} membrane that are close in frequency (see Fig. S3 (c)).
We again employ COMSOL simulations to gain more insight into the combined system (Fig. S1).
The simulations predict a variety of hybridized modes where the Ince-Gaussian type modes expected for the hBN drum are combined with modes of the \ch{Si3N4} membrane \cite{bandresInceGaussianModes2004}. 
This results in a much larger number of modes compared to a bare hBN drum and also affects the rotational orientation of the hBN drum modes.
When taking a closer look at the bottom row in Fig. \ref{fgr:overviewmodes}, it becomes apparent that some of the mode images show motion (bright color) surrounding the typical double maxima of the drum mode, hinting at hybridization with the surrounding \ch{Si3N4} membrane.
As expected, the mode with the lowest Q-factor (second image) shows no motion in the surrounding \ch{Si3N4} membrane, while the one with the highest Q-factor (last image) shows strong motion that even merges with the mode shape of the hBN drum.
The fundamental mode of the hBN drum is not expected to show any hybridization (inset Fig. \ref{fgr:QandMass} (a)), since it sits in a frequency gap in the mode spectrum just after the fundamental mode of the \ch{Si3N4} membrane (arrows in Fig. \ref{fgr:overviewmodes}, Fig. S3).
Instead, for higher order modes of the hBN drum we expect hybridization with modes of the \ch{Si3N4} membrane based on the simulations.

To further investigate this hybridization, we extract $m^{*}$ and Q for all the modes observed in the thermal spectrum.
We also performed measurements both at liquid nitrogen (\SI{77}{K}) and liquid helium (\SI{4}{K}) temperatures. 
The sample was cooled in a liquid helium bath cryostat (Cryomagnetics) and measurements were performed using a fiber based microscope as described in Ref.\citenum{hogeleFiberbasedConfocalMicroscope2008a}.

In Fig. \ref{fgr:QandMass} (b) we show Q as a function of the resonance frequency. 
For a small frequency interval there are often many modes at different Q’s that almost appear in a vertical line, representing groups of modes as the one shown at the bottom of Fig. \ref{fgr:overviewmodes}.
The room-temperature qualtiy factor $Q=\num{1.8e5}$ for the \ch{Si3N4} membrane (see Fig. S3 (a)) is much higher than the one for the fundamental mode of the hBN drum (Fig. \ref{fgr:QandMass} (a)). 
Among the observed thermal spectra whose mode images match modes expected for the hBN drum (\latin{e.g} Fig. \ref{fgr:overviewmodes} bottom row), we have found Q-factors in excess of \num{1e5}, almost reaching the value of the \ch{Si3N4} membrane.

If the motion that we detect on the hBN drum were coupled to motion in the \ch{Si3N4}, we would expect an increase in $m^{*}$ of this hybridized mode, since the \ch{Si3N4} membrane is a much heavier oscillator.
As can be seen in Fig. \ref{fgr:QandMass} (c), where we look at $m^{*}$ instead, we indeed find an increase in these values.
The observed values for room temperature (red markers) fall between a lower limit (red dashed line) corresponding to $m^{*}_{th}$ of the fundamental mode of the hBN drum and an upper limit (green dashed line) corresponding to $m^{*}_{th}$ of the \ch{Si3N4} membrane.
While these limits are only estimates, especially for the circular hBN drum where the effective mass depends on the mode in question \cite{hauerGeneralProcedureThermomechanical2013a}, the behaviour matches our expectations for different degrees of hybridizations.
If we plot the observed Q vs. $m^{*}$ (Fig. \ref{fgr:QandMass} (d)), we find that the modes with higher Q tend to have a higher $m^{*}$, the two values appear to be correlated. 
From this observation we can conclude that, via hybridization, the \ch{Si3N4} membrane can lend its high Q-factor to the hBN at the cost of a higher effective mass.
To explain the shift towards lower $m^{*}$ at lower temperatures, we need to take changes in the resonator geometry during cool-down into account.

\begin{figure}[ht]
  \includegraphics[width=0.5\textwidth]{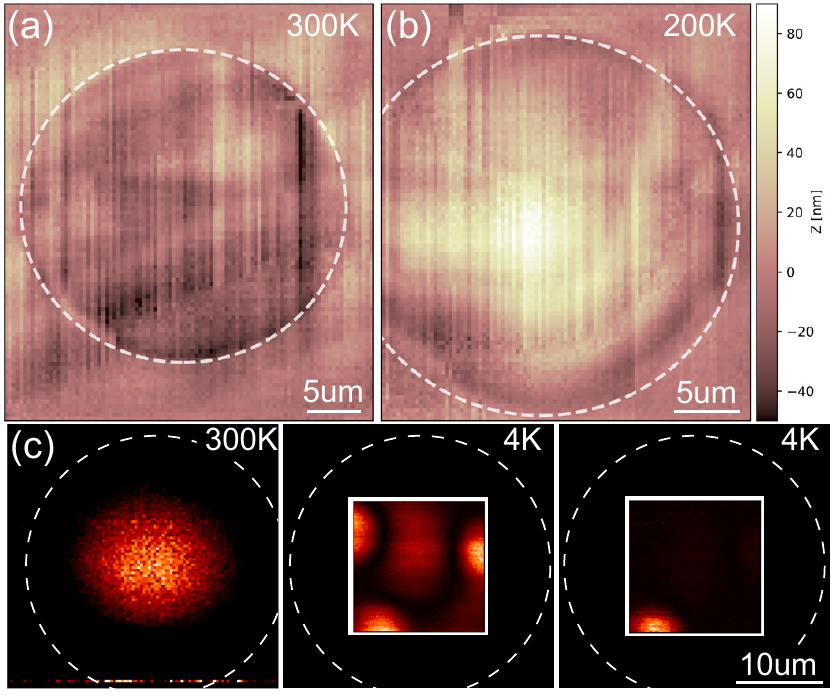}
  \caption{(a),(b) Maps of the height differences extracted from the reflected intensity, showing the bulging effect at lower temperatures. (c) Fundamental hBN mode at room temperature (left), modes of the bulged drum at \SI{4}{K} (middle, right). The image in the middle shows a complex mode that spans a large area, for which thermal motion could not be observed. The image on the right is a more typical mode representing the ones that we were able to characterize (see Fig. \ref{fgr:QandMass}). Dashed circles show the drum edge while the white squares in the last two images indicate the limited scan window at \SI{4}{K}.}
  \label{fgr:LTGraph}
\end{figure}

Bulk hBN is known to have a negative thermal expansion coefficient \cite{paszkowiczLatticeParametersAnisotropic2002}, so, as in graphene, it is reasonable to expect that this property is preserved or even enhanced in the 2D limit \cite{yoonNegativeThermalExpansion2011}.
Therefore, we expect a bulging of the membranes going to low temperatures. 
While no such effect was observed for monolayer hBN \cite{cartamil-buenoMechanicalCharacterizationCleaning2017}, the sample presented here is not thin enough to be clearly in the membrane regime, where the drum would be expected to stay under tension by adhering to the side walls of the underlying substrate \cite{zhengHexagonalBoronNitride2017}.
Indeed, we observe buckling for our sample, as can be seen in the reflected intensity maps in Fig. \ref{fgr:LTGraph} (a) and (b).
While we observe an increase in Q (Fig. \ref{fgr:QandMass}), especially when comparing similar effective masses, the effect is not as big as in other published work \cite{morellHighQualityFactor2016,zandeLargeScaleArraysSingleLayer2010,chenPerformanceMonolayerGraphene2009,
cartamil-buenoMechanicalCharacterizationCleaning2017}.
It is difficult to draw clear comparisons since the nature of the mechanical modes significantly changes after the bulging takes place (see Fig. \ref{fgr:LTGraph} (c)). 
For example, we can not observe the evolution of the Q-factor of the fundamental mode as a function of temperature, since no corresponding mode exists after buckling.

Looking at Fig. \ref{fgr:QandMass}, $m^{*}$ tends towards lower values for lower temperatures, even falling well below the expected value for the fundamental mode of hBN.
This is not surprising, as the mode images reveal that a much smaller area of the membrane takes part in the motion for some of the observed modes in the bulged membranes (Fig. \ref{fgr:LTGraph} (c)).
Unlike at room temperature, we do not observe $m^{*}$ reaching the upper limit given by the motional mass of the \ch{Si3N4} membrane.
We believe that this is due to our inability to observe thermal motion for such modes. 
The thermal motion scales with T and inversely with $m^{*}$ and $\omega$ (Eq.~\ref{eqn:effmass}), limiting the range of observable spectra at low temperatures.

To summarize, the device presented here shows excellent mechanical properties and exhibits mode shapes that are consistent with theory up to high mode orders.
We believe that the most likely explanation for this good performance is the fabrication procedure.
The Si/\ch{SiO2} substrate enables good visibility of any imperfections in the flake and the wet transfer allows the flakes to gently settle onto the \ch{Si3N4} membrane, avoiding inhomogeneous stress in the drum.
While the thickness does not appear to have an effect on the Q of similar devices \cite{zhengHexagonalBoronNitride2017,leeHighFrequencyMoS22013,shandilyaHexagonalBoronNitride2019,
bunchElectromechanicalResonatorsGraphene2007,castellanos-gomezandresSingleLayerMoS22013}, there may be a positive effect due to the large drum diameter of our devices \cite{bartonHighSizeDependentQuality2011}.
Another factor that is known to influence Q is the tension \cite{zandeLargeScaleArraysSingleLayer2010}, which we estimate in our samples to not be any higher compared to similar published work (see supplementary information) \cite{chenPerformanceMonolayerGraphene2009,
cartamil-buenoMechanicalCharacterizationCleaning2017,zhengHexagonalBoronNitride2017}.

The hybridization between the \ch{Si3N4} membrane and the hBN drum not only boosts the Q of the hBN modes, it also allows the selection of different values of Q ($m^{*}$), because the hybridization splits the expected hBN modes into several copies with different degrees of hybridization.
With the addition of temperature control it should be possible to tune this hybridization in a similar way as in Ref.\citenum{schwarzDeviationNormalMode2016,
singhGiantTunableMechanical2020}.

While the gain in Q does not outweigh the increase in $m^{*}$ when estimating the sensitivity of our devices (Fig. S2), \ch{Si3N4} membranes with higher Q (or lower $m^{*}$) have been demonstrated and could make this trade-off more favorable \cite{tsaturyanUltracoherentNanomechanicalResonators2017a}.
This could give access to mechanical modes with outstanding properties, combined with the many unique features of 2D materials.

\begin{acknowledgement}

  We thank Sascha Martin and his team in the machine shop of the
  Physics Department at the University of Basel for help in building
  the apparatus. We acknowledge the support of the Canton Aargau and
  the Swiss National Science Foundation (SNSF) under Ambizione Grant
  No. PZ00P2-161284/1 and Project Grant No. 200020-178863 and via the
  NCCR Quantum Science and Technology (QSIT).

\end{acknowledgement}




\providecommand{\latin}[1]{#1}
\makeatletter
\providecommand{\doi}
  {\begingroup\let\do\@makeother\dospecials
  \catcode`\{=1 \catcode`\}=2 \doi@aux}
\providecommand{\doi@aux}[1]{\endgroup\texttt{#1}}
\makeatother
\providecommand*\mcitethebibliography{\thebibliography}
\csname @ifundefined\endcsname{endmcitethebibliography}
  {\let\endmcitethebibliography\endthebibliography}{}


\begin{mcitethebibliography}{45}
\providecommand*\natexlab[1]{#1}
\providecommand*\mciteSetBstSublistMode[1]{}
\providecommand*\mciteSetBstMaxWidthForm[2]{}
\providecommand*\mciteBstWouldAddEndPuncttrue
  {\def\EndOfBibitem{\unskip.}}
\providecommand*\mciteBstWouldAddEndPunctfalse
  {\let\EndOfBibitem\relax}
\providecommand*\mciteSetBstMidEndSepPunct[3]{}
\providecommand*\mciteSetBstSublistLabelBeginEnd[3]{}
\providecommand*\EndOfBibitem{}
\mciteSetBstSublistMode{f}
\mciteSetBstMaxWidthForm{subitem}{(\alph{mcitesubitemcount})}
\mciteSetBstSublistLabelBeginEnd
  {\mcitemaxwidthsubitemform\space}
  {\relax}
  {\relax}

\bibitem[Bunch \latin{et~al.}(2007)Bunch, van~der Zande, Verbridge, Frank,
  Tanenbaum, Parpia, Craighead, and
  McEuen]{bunchElectromechanicalResonatorsGraphene2007}
Bunch,~J.~S.; van~der Zande,~A.~M.; Verbridge,~S.~S.; Frank,~I.~W.;
  Tanenbaum,~D.~M.; Parpia,~J.~M.; Craighead,~H.~G.; McEuen,~P.~L.
  Electromechanical {{Resonators}} from {{Graphene Sheets}}. \emph{Science}
  \textbf{2007}, \emph{315}, 490--493
\bibitem[Zheng \latin{et~al.}(2017)Zheng, Lee, and
  Feng]{zhengHexagonalBoronNitride2017}
Zheng,~X.-Q.; Lee,~J.; Feng,~P. X.-L. Hexagonal Boron Nitride Nanomechanical
  Resonators with Spatially Visualized Motion. \emph{Microsystems \&
  Nanoengineering} \textbf{2017}, \emph{3}, 17038
\bibitem[{Castellanos-Gomez Andres} \latin{et~al.}(2013){Castellanos-Gomez
  Andres}, {van Leeuwen Ronald}, {Buscema Michele}, {van der Zant Herre S. J.},
  {Steele Gary A.}, and {Venstra Warner
  J.}]{castellanos-gomezandresSingleLayerMoS22013}
{Castellanos-Gomez Andres},; {van Leeuwen Ronald},; {Buscema Michele},; {van
  der Zant Herre S. J.},; {Steele Gary A.},; {Venstra Warner J.},
  Single-{{Layer MoS2 Mechanical Resonators}}. \emph{Advanced Materials}
  \textbf{2013}, \emph{25}, 6719--6723
\bibitem[Steeneken \latin{et~al.}(2021)Steeneken, Dolleman, Davidovikj,
  Alijani, and {van der Zant}]{steenekenDynamics2DMaterial2021}
Steeneken,~P.~G.; Dolleman,~R.~J.; Davidovikj,~D.; Alijani,~F.; {van der
  Zant},~H. S.~J. Dynamics of {{2D Material Membranes}}. \emph{2D Materials}
  \textbf{2021}, \emph{8}, 042001
\bibitem[Lee \latin{et~al.}(2008)Lee, Wei, Kysar, and
  Hone]{leeMeasurementElasticProperties2008}
Lee,~C.; Wei,~X.; Kysar,~J.~W.; Hone,~J. Measurement of the {{Elastic
  Properties}} and {{Intrinsic Strength}} of {{Monolayer Graphene}}.
  \emph{Science} \textbf{2008}, \emph{321}, 385--388
\bibitem[Chen \latin{et~al.}(2009)Chen, Rosenblatt, Bolotin, Kalb, Kim,
  Kymissis, Stormer, Heinz, and Hone]{chenPerformanceMonolayerGraphene2009}
Chen,~C.; Rosenblatt,~S.; Bolotin,~K.~I.; Kalb,~W.; Kim,~P.; Kymissis,~I.;
  Stormer,~H.~L.; Heinz,~T.~F.; Hone,~J. Performance of Monolayer Graphene
  Nanomechanical Resonators with Electrical Readout. \emph{Nature
  Nanotechnology} \textbf{2009}, \emph{4}, 861--867
\bibitem[Chen \latin{et~al.}(2013)Chen, Lee, Deshpande, Lee, Lekas, Shepard,
  and Hone]{chenGrapheneMechanicalOscillators2013}
Chen,~C.; Lee,~S.; Deshpande,~V.~V.; Lee,~G.-H.; Lekas,~M.; Shepard,~K.;
  Hone,~J. Graphene Mechanical Oscillators with Tunable Frequency. \emph{Nature
  Nanotechnology} \textbf{2013}, \emph{8}, 923--927
\bibitem[Gong \latin{et~al.}(2017)Gong, Li, Li, Ji, Stern, Xia, Cao, Bao, Wang,
  Wang, Qiu, Cava, Louie, Xia, and
  Zhang]{gongDiscoveryIntrinsicFerromagnetism2017a}
Gong,~C.; Li,~L.; Li,~Z.; Ji,~H.; Stern,~A.; Xia,~Y.; Cao,~T.; Bao,~W.;
  Wang,~C.; Wang,~Y.; Qiu,~Z.~Q.; Cava,~R.~J.; Louie,~S.~G.; Xia,~J.; Zhang,~X.
  Discovery of Intrinsic Ferromagnetism in Two-Dimensional van Der {{Waals}}
  Crystals. \emph{Nature} \textbf{2017}, \emph{546}, 265--269
\bibitem[Huang \latin{et~al.}(2017)Huang, Clark, {Navarro-Moratalla}, Klein,
  Cheng, Seyler, Zhong, Schmidgall, McGuire, Cobden, Yao, Xiao,
  {Jarillo-Herrero}, and Xu]{huangLayerdependentFerromagnetismVan2017a}
Huang,~B.; Clark,~G.; {Navarro-Moratalla},~E.; Klein,~D.~R.; Cheng,~R.;
  Seyler,~K.~L.; Zhong,~D.; Schmidgall,~E.; McGuire,~M.~A.; Cobden,~D.~H.;
  Yao,~W.; Xiao,~D.; {Jarillo-Herrero},~P.; Xu,~X. Layer-Dependent
  Ferromagnetism in a van Der {{Waals}} Crystal down to the Monolayer Limit.
  \emph{Nature} \textbf{2017}, \emph{546}, 270--273
\bibitem[Splendiani \latin{et~al.}(2010)Splendiani, Sun, Zhang, Li, Kim, Chim,
  Galli, and Wang]{splendianiEmergingPhotoluminescenceMonolayer2010}
Splendiani,~A.; Sun,~L.; Zhang,~Y.; Li,~T.; Kim,~J.; Chim,~C.-Y.; Galli,~G.;
  Wang,~F. Emerging {{Photoluminescence}} in {{Monolayer MoS2}}. \emph{Nano
  Letters} \textbf{2010}, \emph{10}, 1271--1275
\bibitem[Srivastava \latin{et~al.}(2015)Srivastava, Sidler, Allain, Lembke,
  Kis, and Imamo{\u g}lu]{srivastavaOpticallyActiveQuantum2015}
Srivastava,~A.; Sidler,~M.; Allain,~A.~V.; Lembke,~D.~S.; Kis,~A.; Imamo{\u
  g}lu,~A. Optically Active Quantum Dots in Monolayer {{WSe2}}. \emph{Nature
  Nanotechnology} \textbf{2015}, \emph{10}, 491--496
\bibitem[Novoselov \latin{et~al.}(2016)Novoselov, Mishchenko, Carvalho, and
  Castro~Neto]{novoselov2DMaterialsVan2016a}
Novoselov,~K.~S.; Mishchenko,~A.; Carvalho,~A.; Castro~Neto,~A.~H. {{2D}}
  Materials and van Der {{Waals}} Heterostructures. \emph{Science}
  \textbf{2016}, \emph{353}, aac9439
\bibitem[Falin \latin{et~al.}(2017)Falin, Cai, Santos, Scullion, Qian, Zhang,
  Yang, Huang, Watanabe, Taniguchi, Barnett, Chen, Ruoff, and
  Li]{falinMechanicalPropertiesAtomically2017}
Falin,~A.; Cai,~Q.; Santos,~E. J.~G.; Scullion,~D.; Qian,~D.; Zhang,~R.;
  Yang,~Z.; Huang,~S.; Watanabe,~K.; Taniguchi,~T.; Barnett,~M.~R.; Chen,~Y.;
  Ruoff,~R.~S.; Li,~L.~H. Mechanical Properties of Atomically Thin Boron
  Nitride and the Role of Interlayer Interactions. \emph{Nature Communications}
  \textbf{2017}, \emph{8}, 15815
\bibitem[Watanabe \latin{et~al.}(2004)Watanabe, Taniguchi, and
  Kanda]{watanabeDirectbandgapPropertiesEvidence2004}
Watanabe,~K.; Taniguchi,~T.; Kanda,~H. Direct-Bandgap Properties and Evidence
  for Ultraviolet Lasing of Hexagonal Boron Nitride Single Crystal.
  \emph{Nature Materials} \textbf{2004}, \emph{3}, 404--409
\bibitem[Lee \latin{et~al.}(2019)Lee, Jeong, Jung, and
  Yee]{leeRefractiveIndexDispersion2019}
Lee,~S.-Y.; Jeong,~T.-Y.; Jung,~S.; Yee,~K.-J. Refractive {{Index Dispersion}}
  of {{Hexagonal Boron Nitride}} in the {{Visible}} and {{Near-Infrared}}.
  \emph{physica status solidi (b)} \textbf{2019}, \emph{256}, 1800417
\bibitem[Tran \latin{et~al.}(2016)Tran, Bray, Ford, Toth, and
  Aharonovich]{tranQuantumEmissionHexagonal2016}
Tran,~T.~T.; Bray,~K.; Ford,~M.~J.; Toth,~M.; Aharonovich,~I. Quantum Emission
  from Hexagonal Boron Nitride Monolayers. \emph{Nature Nanotechnology}
  \textbf{2016}, \emph{11}, 37--41
\bibitem[Mendelson \latin{et~al.}(2020)Mendelson, Doherty, Toth, Aharonovich,
  and Tran]{mendelsonStrainInducedModificationOptical2020}
Mendelson,~N.; Doherty,~M.; Toth,~M.; Aharonovich,~I.; Tran,~T.~T.
  Strain-{{Induced Modification}} of the {{Optical Characteristics}} of
  {{Quantum Emitters}} in {{Hexagonal Boron Nitride}}. \emph{Advanced
  Materials} \textbf{2020}, \emph{32}, 1908316
\bibitem[Lazi{\'c} \latin{et~al.}(2019)Lazi{\'c}, Espinha, Yanguas, Gibaja,
  Zamora, Ares, Chhowalla, Paz, Burgos, {Hern{\'a}ndez-M{\'i}nguez}, Santos,
  and van~der Meulen]{lazicDynamicallyTunedNonclassical2019}
Lazi{\'c},~S.; Espinha,~A.; Yanguas,~S.~P.; Gibaja,~C.; Zamora,~F.; Ares,~P.;
  Chhowalla,~M.; Paz,~W.~S.; Burgos,~J. J.~P.; {Hern{\'a}ndez-M{\'i}nguez},~A.;
  Santos,~P.~V.; van~der Meulen,~H.~P. Dynamically Tuned Non-Classical Light
  Emission from Atomic Defects in Hexagonal Boron Nitride. \emph{Communications
    Physics} \textbf{2019}, \emph{2}, 1--8
\bibitem[Shandilya \latin{et~al.}(2019)Shandilya, Fr{\"o}ch, Mitchell, Lake,
  Kim, Toth, Behera, Healey, Aharonovich, and
  Barclay]{shandilyaHexagonalBoronNitride2019}
Shandilya,~P.~K.; Fr{\"o}ch,~J.~E.; Mitchell,~M.; Lake,~D.~P.; Kim,~S.;
  Toth,~M.; Behera,~B.; Healey,~C.; Aharonovich,~I.; Barclay,~P.~E. Hexagonal
  {{Boron Nitride Cavity Optomechanics}}. \emph{Nano Letters} \textbf{2019},
  \emph{19}, 1343--1350
\bibitem[Lee \latin{et~al.}(2013)Lee, Wang, He, Shan, and
  Feng]{leeHighFrequencyMoS22013}
Lee,~J.; Wang,~Z.; He,~K.; Shan,~J.; Feng,~P. X.-L. High {{Frequency MoS2
  Nanomechanical Resonators}}. \emph{ACS Nano} \textbf{2013}, \emph{7},
  6086--6091
\bibitem[{Castellanos-Gomez} \latin{et~al.}(2014){Castellanos-Gomez}, Buscema,
  Molenaar, Singh, Janssen, van~der Zant, and
  Steele]{castellanos-gomezDeterministicTransferTwodimensional2014}
{Castellanos-Gomez},~A.; Buscema,~M.; Molenaar,~R.; Singh,~V.; Janssen,~L.;
  van~der Zant,~H. S.~J.; Steele,~G.~A. Deterministic Transfer of
  Two-Dimensional Materials by All-Dry Viscoelastic Stamping. \emph{2D
  Materials} \textbf{2014}, \emph{1}, 011002
\bibitem[Ruelle \latin{et~al.}(2022)Ruelle, Jaeger, Fogliano, Braakman, and
  Poggio]{ruelleTunableFiberFabry2022}
Ruelle,~T.; Jaeger,~D.; Fogliano,~F.; Braakman,~F.; Poggio,~M. A Tunable Fiber
  {{Fabry}}\textendash{{Perot}} Cavity for Hybrid Optomechanics Stabilized at 4
  {{K}}. \emph{Review of Scientific Instruments} \textbf{2022}, \emph{93},
  095003
\bibitem[{Flowers-Jacobs} \latin{et~al.}(2012){Flowers-Jacobs}, Hoch, Sankey,
  Kashkanova, Jayich, Deutsch, Reichel, and
  Harris]{flowers-jacobsFibercavitybasedOptomechanicalDevice2012a}
{Flowers-Jacobs},~N.~E.; Hoch,~S.~W.; Sankey,~J.~C.; Kashkanova,~A.;
  Jayich,~A.~M.; Deutsch,~C.; Reichel,~J.; Harris,~J. G.~E. Fiber-Cavity-Based
  Optomechanical Device. \emph{Applied Physics Letters} \textbf{2012},
  \emph{101}, 221109
\bibitem[Rochau \latin{et~al.}(2021)Rochau, S{\'a}nchez~Arribas, Brieussel,
  Stapfner, Hunger, and Weig]{rochauDynamicalBackactionUltrahighFinesse2021a}
Rochau,~F.; S{\'a}nchez~Arribas,~I.; Brieussel,~A.; Stapfner,~S.; Hunger,~D.;
  Weig,~E.~M. Dynamical {{Backaction}} in an {{Ultrahigh-Finesse Fiber-Based
  Microcavity}}. \emph{Physical Review Applied} \textbf{2021}, \emph{16},
  014013
\bibitem[Davidovikj \latin{et~al.}(2016)Davidovikj, Slim, Bueno, {van der
  Zant}, Steeneken, and Venstra]{davidovikjVisualizingMotionGraphene2016}
Davidovikj,~D.; Slim,~J.~J.; Bueno,~S. J.~C.; {van der Zant},~H. S.~J.;
  Steeneken,~P.~G.; Venstra,~W.~J. Visualizing the Motion of Graphene
  Nanodrums. \emph{Nano Letters} \textbf{2016}, \emph{16}, 2768--2773
\bibitem[{Garcia-Sanchez} \latin{et~al.}(2008){Garcia-Sanchez}, {van der
  Zande}, Paulo, Lassagne, McEuen, and
  Bachtold]{garcia-sanchezImagingMechanicalVibrations2008}
{Garcia-Sanchez},~D.; {van der Zande},~A.~M.; Paulo,~A.~S.; Lassagne,~B.;
  McEuen,~P.~L.; Bachtold,~A. Imaging {{Mechanical Vibrations}} in {{Suspended
  Graphene Sheets}}. \emph{Nano Letters} \textbf{2008}, \emph{8},
  1399--1403
\bibitem[Reina \latin{et~al.}(2009)Reina, Jia, Ho, Nezich, Son, Bulovic,
  Dresselhaus, and Kong]{reinaLargeAreaFewLayer2009}
Reina,~A.; Jia,~X.; Ho,~J.; Nezich,~D.; Son,~H.; Bulovic,~V.;
  Dresselhaus,~M.~S.; Kong,~J. Large {{Area}}, {{Few-Layer Graphene Films}} on
  {{Arbitrary Substrates}} by {{Chemical Vapor Deposition}}. \emph{Nano
  Letters} \textbf{2009}, \emph{9}, 30--35
\bibitem[van~der Zande \latin{et~al.}(2010)van~der Zande, Barton, Alden,
  {Ruiz-Vargas}, Whitney, Pham, Park, Parpia, Craighead, and
  McEuen]{zandeLargeScaleArraysSingleLayer2010}
van~der Zande,~A.~M.; Barton,~R.~A.; Alden,~J.~S.; {Ruiz-Vargas},~C.~S.;
  Whitney,~W.~S.; Pham,~P. H.~Q.; Park,~J.; Parpia,~J.~M.; Craighead,~H.~G.;
  McEuen,~P.~L. Large-{{Scale Arrays}} of {{Single-Layer Graphene Resonators}}.
  \emph{Nano Letters} \textbf{2010}, \emph{10}, 4869--4873
\bibitem[{Cartamil-Bueno} \latin{et~al.}(2017){Cartamil-Bueno}, Cavalieri,
  Wang, Houri, Hofmann, and van~der
  Zant]{cartamil-buenoMechanicalCharacterizationCleaning2017}
{Cartamil-Bueno},~S.~J.; Cavalieri,~M.; Wang,~R.; Houri,~S.; Hofmann,~S.;
  van~der Zant,~H. S.~J. Mechanical Characterization and Cleaning of {{CVD}}
  Single-Layer h-{{BN}} Resonators. \emph{npj 2D Materials and Applications}
  \textbf{2017}, \emph{1}, 16
\bibitem[Schwarz \latin{et~al.}(2016)Schwarz, Pigeau, {Mercier de L{\'e}pinay},
  Kuhn, Kalita, Bendiab, Marty, Bouchiat, and
  Arcizet]{schwarzDeviationNormalMode2016}
Schwarz,~C.; Pigeau,~B.; {Mercier de L{\'e}pinay},~L.; Kuhn,~A.~G.; Kalita,~D.;
  Bendiab,~N.; Marty,~L.; Bouchiat,~V.; Arcizet,~O. Deviation from the {{Normal
  Mode Expansion}} in a {{Coupled Graphene-Nanomechanical System}}.
  \emph{Physical Review Applied} \textbf{2016}, \emph{6}, 064021
\bibitem[Singh \latin{et~al.}(2020)Singh, Sarkar, Guria, Nicholl, Chakraborty,
  Bolotin, and Ghosh]{singhGiantTunableMechanical2020}
Singh,~R.; Sarkar,~A.; Guria,~C.; Nicholl,~R.~J.; Chakraborty,~S.;
  Bolotin,~K.~I.; Ghosh,~S. Giant {{Tunable Mechanical Nonlinearity}} in
  {{Graphene}}\textendash{{Silicon Nitride Hybrid Resonator}}. \emph{Nano
  Letters} \textbf{2020}, \emph{20}, 4659--4666
\bibitem[Reiche \latin{et~al.}(2015)Reiche, K{\"o}rner, B{\"u}chner, and
  M{\"u}hl]{reicheIntroductionCoresonantDetection2015}
Reiche,~C.~F.; K{\"o}rner,~J.; B{\"u}chner,~B.; M{\"u}hl,~T. Introduction of a
  Co-Resonant Detection Concept for Mechanical Oscillation-Based Sensors.
  \emph{Nanotechnology} \textbf{2015}, \emph{26}, 335501
\bibitem[Zheng \latin{et~al.}(2015)Zheng, Lee, and
  Feng]{zhengHexagonalBoronNitride2015}
Zheng,~X.-Q.; Lee,~J.; Feng,~P. X.-L. Hexagonal Boron Nitride (h-{{BN}})
  Nanomechanical Resonators with Temperature-Dependent Multimode Operations.
  2015 {{Transducers}} - 2015 18th {{International Conference}} on
  {{Solid-State Sensors}}, {{Actuators}} and {{Microsystems}}
  ({{TRANSDUCERS}}). 2015; pp 1393--1396
\bibitem[Barg \latin{et~al.}(2017)Barg, Tsaturyan, Belhage, Nielsen, M{\o}ller,
  and Schliesser]{bargMeasuringImagingNanomechanical2017}
Barg,~A.; Tsaturyan,~Y.; Belhage,~E.; Nielsen,~W. H.~P.; M{\o}ller,~C.~B.;
  Schliesser,~A. Measuring and Imaging Nanomechanical Motion with Laser Light.
  \emph{Applied Physics B} \textbf{2017}, \emph{123}, 8
\bibitem[Hauer \latin{et~al.}(2013)Hauer, Doolin, Beach, and
  Davis]{hauerGeneralProcedureThermomechanical2013a}
Hauer,~B.~D.; Doolin,~C.; Beach,~K. S.~D.; Davis,~J.~P. A General Procedure for
  Thermomechanical Calibration of Nano/Micro-Mechanical Resonators.
  \emph{Annals of Physics} \textbf{2013}, \emph{339}, 181--207
\bibitem[{van Leeuwen} \latin{et~al.}(2014){van Leeuwen}, {Castellanos-Gomez},
  Steele, {van der Zant}, and
  Venstra]{vanleeuwenTimedomainResponseAtomically2014}
{van Leeuwen},~R.; {Castellanos-Gomez},~A.; Steele,~G.~A.; {van der Zant},~H.
  S.~J.; Venstra,~W.~J. Time-Domain Response of Atomically Thin {{MoS2}}
  Nanomechanical Resonators. \emph{Applied Physics Letters} \textbf{2014},
  \emph{105}, 041911
\bibitem[Kramer \latin{et~al.}(2015)Kramer, {van Dorp}, {van Leeuwen}, and
  Venstra]{kramerStraindependentDampingNanomechanical2015}
Kramer,~E.; {van Dorp},~J.; {van Leeuwen},~R.; Venstra,~W.~J. Strain-Dependent
  Damping in Nanomechanical Resonators from Thin {{MoS2}} Crystals.
  \emph{Applied Physics Letters} \textbf{2015}, \emph{107}, 091903
\bibitem[Bandres and {Guti{\'e}rrez-Vega}(2004)Bandres, and
  {Guti{\'e}rrez-Vega}]{bandresInceGaussianModes2004}
Bandres,~M.~A.; {Guti{\'e}rrez-Vega},~J.~C. Ince\textendash{{Gaussian}} Modes
  of the Paraxial Wave Equation and Stable Resonators. \emph{Journal of the
  Optical Society of America A} \textbf{2004}, \emph{21}, 873
\bibitem[H{\"o}gele \latin{et~al.}(2008)H{\"o}gele, Seidl, Kroner, Karrai,
  Schulhauser, Sqalli, Scrimgeour, and
  Warburton]{hogeleFiberbasedConfocalMicroscope2008a}
H{\"o}gele,~A.; Seidl,~S.; Kroner,~M.; Karrai,~K.; Schulhauser,~C.; Sqalli,~O.;
  Scrimgeour,~J.; Warburton,~R.~J. Fiber-Based Confocal Microscope for
  Cryogenic Spectroscopy. \emph{Review of Scientific Instruments}
  \textbf{2008}, \emph{79}, 023709
\bibitem[Paszkowicz \latin{et~al.}(2002)Paszkowicz, Pelka, Knapp, Szyszko, and
  Podsiadlo]{paszkowiczLatticeParametersAnisotropic2002}
Paszkowicz,~W.; Pelka,~J.; Knapp,~M.; Szyszko,~T.; Podsiadlo,~S. Lattice
  Parameters and Anisotropic Thermal Expansion of Hexagonal Boron Nitride in
  the 10\textendash 297.5~{{K}} Temperature Range. \emph{Applied Physics A}
  \textbf{2002}, \emph{75}, 431--435
\bibitem[Yoon \latin{et~al.}(2011)Yoon, Son, and
  Cheong]{yoonNegativeThermalExpansion2011}
Yoon,~D.; Son,~Y.-W.; Cheong,~H. Negative {{Thermal Expansion Coefficient}} of
  {{Graphene Measured}} by {{Raman Spectroscopy}}. \emph{Nano Letters}
  \textbf{2011}, \emph{11}, 3227--3231
\bibitem[Morell \latin{et~al.}(2016)Morell, {Reserbat-Plantey}, Tsioutsios,
  Sch{\"a}dler, Dubin, Koppens, and Bachtold]{morellHighQualityFactor2016}
Morell,~N.; {Reserbat-Plantey},~A.; Tsioutsios,~I.; Sch{\"a}dler,~K.~G.;
  Dubin,~F.; Koppens,~F. H.~L.; Bachtold,~A. High {{Quality Factor Mechanical
  Resonators Based}} on {{WSe2 Monolayers}}. \emph{Nano Letters} \textbf{2016},
  \emph{16}, 5102--5108
\bibitem[Barton \latin{et~al.}(2011)Barton, Ilic, {van der Zande}, Whitney,
  McEuen, Parpia, and Craighead]{bartonHighSizeDependentQuality2011}
Barton,~R.~A.; Ilic,~B.; {van der Zande},~A.~M.; Whitney,~W.~S.; McEuen,~P.~L.;
  Parpia,~J.~M.; Craighead,~H.~G. High, {{Size-Dependent Quality Factor}} in an
  {{Array}} of {{Graphene Mechanical Resonators}}. \emph{Nano Letters}
  \textbf{2011}, \emph{11}, 1232--1236
\bibitem[Tsaturyan \latin{et~al.}(2017)Tsaturyan, Barg, Polzik, and
  Schliesser]{tsaturyanUltracoherentNanomechanicalResonators2017a}
Tsaturyan,~Y.; Barg,~A.; Polzik,~E.~S.; Schliesser,~A. Ultracoherent
  Nanomechanical Resonators via Soft Clamping and Dissipation Dilution.
  \emph{Nature Nanotechnology} \textbf{2017}, \emph{12}, 776--783
\end{mcitethebibliography}
\end{document}


\pagebreak

\section{COMSOL simulations}

\begin{figure}[H]
  \includegraphics[width=\textwidth]{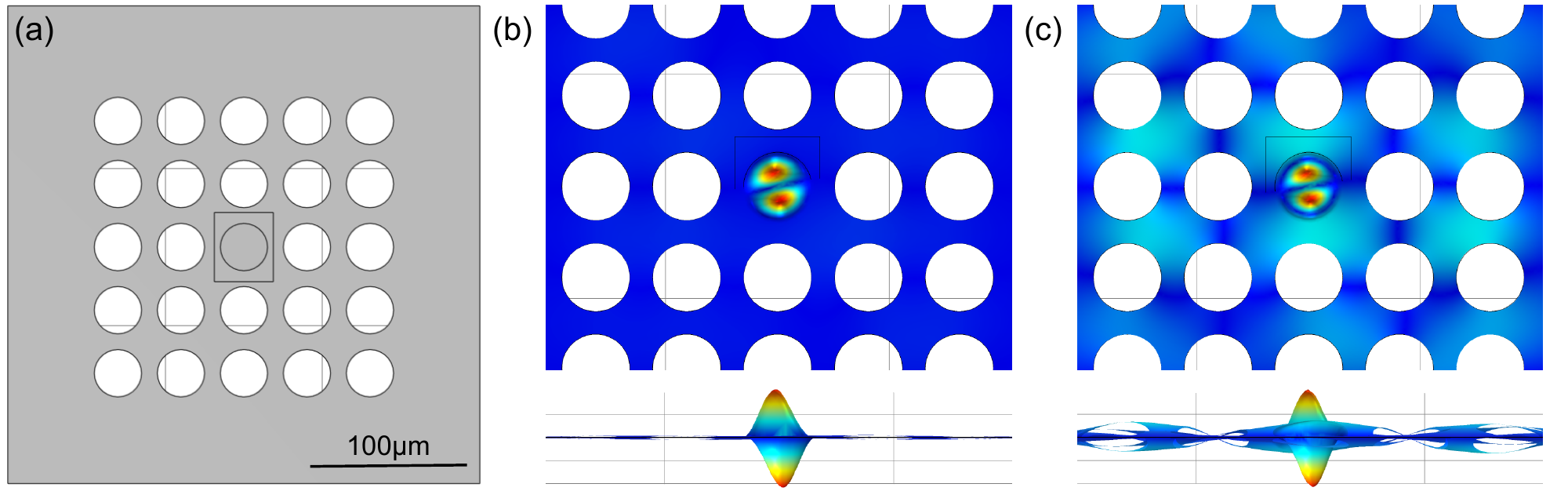}
  \caption{(a) Simplified geometry used for our simulations. (b),(c) Top and side view of two copies of hBN drum modes with different degrees of hybridization with the \ch{Si3N4} membrane.}
  \label{fgr:Comsol}
\end{figure}

For the geometry of the COMSOL simulation we approximate the hBN flake as a rectangle covering the central hole of the \ch{Si3N4} membrane, while the \ch{Si3N4} membrane itself could be accurately reproduced due to its well-defined shape.
The thickness of the membrane is \SI{200}{\nm} and that of the flake is \SI{48}{\nm} in accordance with the AFM measurement shown in Fig.1(c) of the main text. 

The reported values for the Young's modulus of hBN vary greatly \cite{zhengHexagonalBoronNitride2017,cartamil-buenoMechanicalCharacterizationCleaning2017}; we have chosen the average value of \SI{392}{\GPa} reported in Ref.~\citenum{zhengHexagonalBoronNitride2017} for our simulations.
We then match the simulated frequency of the fundamental mode of the hBN drum to the one we observe experimentally using the pre-tension of the flake, resulting in a value of \SI{0.1}{\N/m}.
The stoichiometric \ch{Si3N4} membrane has a tailored stress of \SI{900}{\MPa}.

The mode shapes are evaluated as a shell instead of a membrane to take bending stiffness into account which is expected to play a role given the thickness of our hBN flake \cite{zhengHexagonalBoronNitride2017}. We introduce asymmetry into the system with a 1\% deviation from the square shape of the \ch{Si3N4} membrane and the rectangular shape of the hBN flake.

\pagebreak

\section{Force and mass sensitivity estimations}

\begin{figure}[H]
  \includegraphics[width=\textwidth]{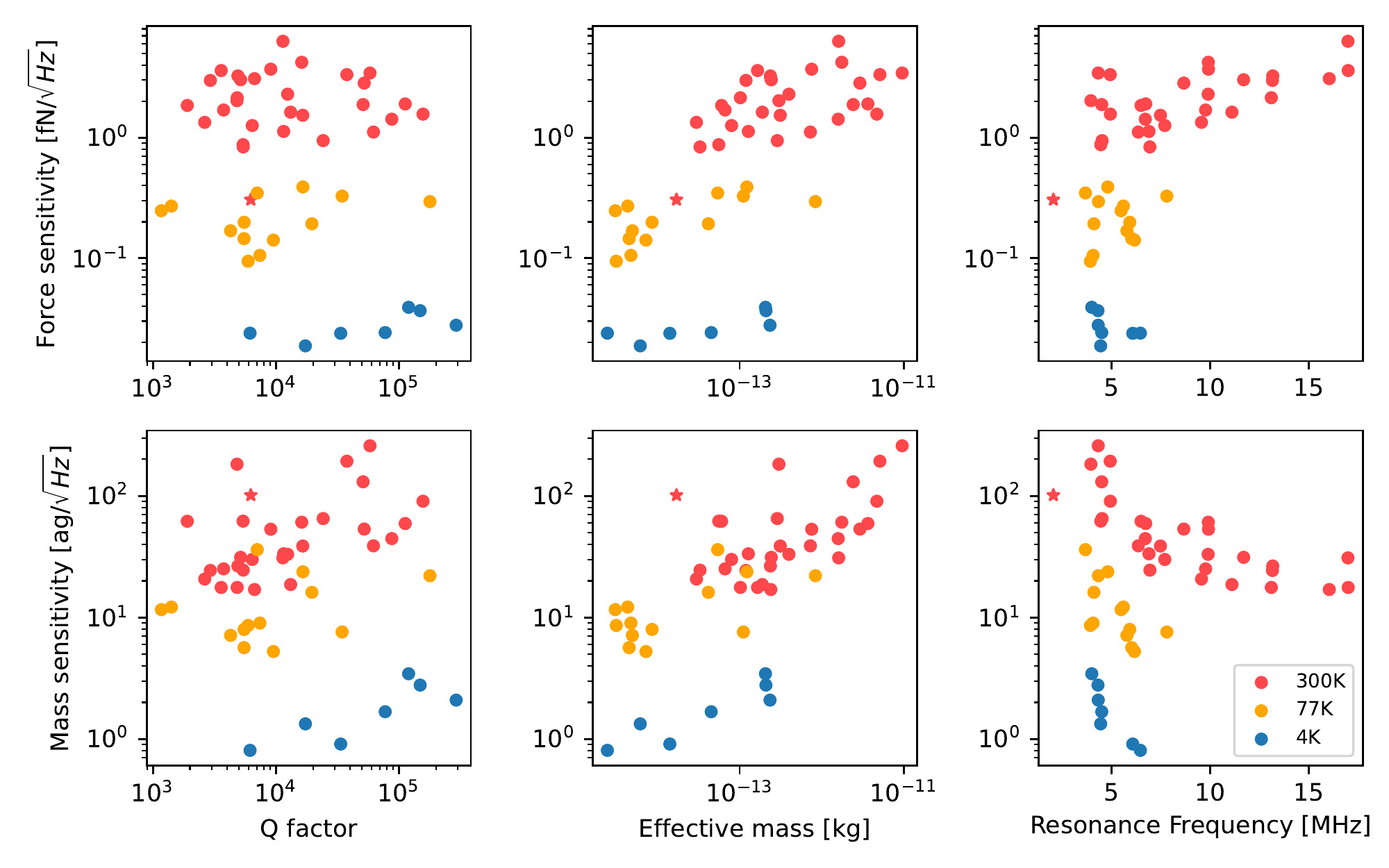}
  \caption{Force sensitivity (mass sensitivity) shown vs. Q, $m^{*}$ and resonance frequency in the top (bottom) row. The star shaped marker shows the fundamental mode of the hBN drum.}
  \label{fgr:Sens}
\end{figure}

The force and mass sensitivities in units of $[N/\sqrt{Hz]}$ and $[kg/\sqrt{Hz}]$ are given by
\begin{equation}
S_{f}^{-1/2}(\omega)=\sqrt{4 k_{B} T \Gamma m^{*}}
	\label{eqn:forcesens}
\end{equation}

\begin{equation}
S_{m}^{-1/2}(\omega)=\sqrt{2 k_{B} T \Gamma m^{*}} \frac{2}{x_{0} \omega_{0}^{2}}
	\label{eqn:masssens}
\end{equation}
with $x_0$ the oscillation amplitude (assumed to be \SI{1}{\nm} for the results shown in Fig. \ref{fgr:Sens}).

In Fig.\ref{fgr:Sens} the resulting sensitivities for the mechanical modes of the hBN drum are plotted. 
For the force sensitivity at room temperature the best values are predicted for the fundamental mode. This indicates that for our system, the gain in Q does not outweigh the increase in $m^{*}$, since the fundamental mode of hBN has the lowest effective mass. 
The mass sensitivity has an additional term $\frac{2}{x_{0} \omega_{0}^{2}}$, giving it a stronger dependency on frequency, but no increase in sensitivity can be observed for the strongly hybridized modes with high values of Q and $m^{*}$.

Going to \SI{77}{\K} and then to \SI{4}{\K} not only improves the sensitivities due to their temperature dependence, but also because of the reduction of $m^{*}$ for the smaller modes in the bulged membrane.

\pagebreak

\section{Mechanical properties of the \ch{Si3N4} membrane}

\begin{figure}[H]
  \includegraphics[width=\textwidth]{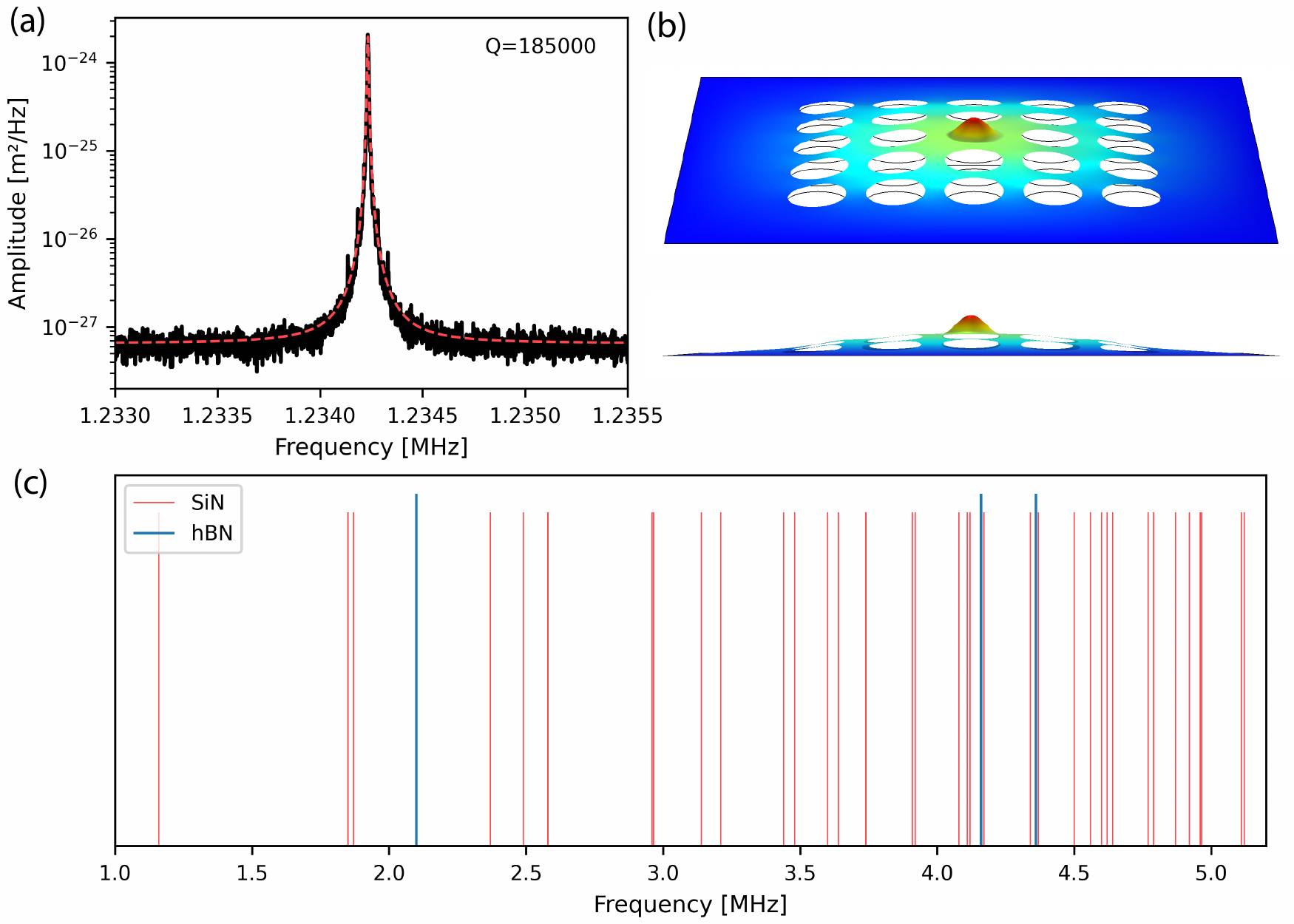}
  \caption{(a) Thermal motion of the fundamental mode of the \ch{Si3N4} membrane. (b) COMSOL simulation of the mode shown in (a). (c) Predicted frequencies for the mechancial modes of the isolated \ch{Si3N4} membrane (orange) and the isolated hBN drum (blue).}
  \label{fgr:SiNmech}
\end{figure}

The fundamental mode of the \ch{Si3N4} membrane has a Q of $\num{1.8e5}$ at room temperature. 
This mode is lower in frequency than the fundamental mode of hBN by almost a factor of two, so no hybridization is expected. 
This shows that the highest Q of hybridized modes shown in the main text, of around $\num{1.5e5}$ is not far from this highest value that we measured for this sample.
In Fig.\ref{fgr:SiNmech}(b) we show the simulated mode shape for the fundamental mode of the \ch{Si3N4} membrane. 
Apart from the expected overall shape, we can observe that the hBN still interacts with this mode by effectively increasing the amplitude due to its different mechanical properties.
In Fig.\ref{fgr:SiNmech}(c) we show how densely packed the frequency spacings become past \SI{3}{\MHz} for the \ch{Si3N4} membrane, explaining how the fundamental mode of the hBN drum is not hybridized, while the higher order modes tend to be hybridized to varying degrees.

\providecommand{\latin}[1]{#1}
\makeatletter
\providecommand{\doi}
  {\begingroup\let\do\@makeother\dospecials
  \catcode`\{=1 \catcode`\}=2 \doi@aux}
\providecommand{\doi@aux}[1]{\endgroup\texttt{#1}}
\makeatother
\providecommand*\mcitethebibliography{\thebibliography}
\csname @ifundefined\endcsname{endmcitethebibliography}
  {\let\endmcitethebibliography\endthebibliography}{}